\documentstyle[aps,epsfig,psfig]{revtex}
\begin{document}

\draft  


\title{Theory of Quantum Annealing of an Ising Spin Glass}

\author{
Giuseppe E. Santoro$^{(1)}$, Roman Marto\v{n}\'ak$^{(2,3)}$, Erio Tosatti$^{(1,4)}$,
and Roberto Car$^{(5)}$ 
}

\address{
$^{(1)}$ International School for Advanced Studies (SISSA) and INFM (UdR SISSA),
Trieste, Italy \\
$^{(2)}$ Swiss Centre for Scientific Computing, Manno, and
ETH-Z\"urich, Physical Chemistry, Z\"urich, Switzerland \\
$^{(3)}$ Dept. of Physics,
  Slovak Technical University (FEI),
  Bratislava, Slovakia\\
$^{(4)}$ International Center for Theoretical Physics (ICTP), P.O.Box 586,
  Trieste, Italy \\
$^{(5)}$ Princeton University, Dept. of Chemistry and Princeton Materials Institute, 
Princeton, USA \\  
}

\date{\today}

\maketitle

\begin{abstract}
Probing the lowest energy configuration of a complex system by quantum 
annealing was recently found to be more effective than its classical, 
thermal counterpart. Comparing classical and quantum Monte Carlo 
annealing protocols on the random two-dimensional Ising model we 
confirm the superiority of quantum annealing relative to classical 
annealing. We also propose a theory of quantum annealing, based on a 
cascade of Landau-Zener tunneling events.  
For both classical and quantum annealing, the residual energy after annealing 
is inversely proportional to a power of the logarithm of the annealing time, 
but the quantum case has a larger power which makes it faster
\end{abstract}

The annealing of disordered and complex systems towards their
optimal (or lowest energy) state is a central problem in statistical physics,
with impact in a large variety of areas.
The unknown ground state of a system can be approximated by slow-rate
cooling of a real or fictitious temperature: the slower the cooling,
the closer the approximation.\cite{Kirk_CA,Cerny}
Although this kind of standard classical annealing (CA) has been extensively 
investigated over the last two decades\cite{Kirk_CA,Cerny,Huse_Fisher},
and is routinely used in a variety of technological applications,
such as chip circuitry design, 
the premium on any alternative better optimization algorithms would 
certainly be enormous.

Recent results of Brooke {\it et al.\/} \cite{aeppli,aeppli2}, on the spin 1/2
disordered Ising ferromagnet LiHo$_{0.44}$Y$_{0.56}$F$_4$ 
suggested however that a different, {\em quantum} annealing (QA) procedure
works surprisingly better than classical annealing.
In QA, temperature is replaced by a quantum mechanical kinetic Hamiltonian term
-- in the specific case a transverse magnetic field $\Gamma$ mixing the up
and down spin states at each site. Initially the quantum
perturbation starts out so large in magnitude as to completely
disorder the system even at zero temperature. When the transverse
field is subsequently reduced to zero at some slow rate $1/\tau$,
the system is ``annealed'' towards its ground state, much in
the same way as when its temperature is reduced to zero in CA.
The question is which of the two, CA or QA, works better, and how and why.
Experimental comparison of the properties displayed by the same system
transported from the same initial state A -- a classical high-T state --
to the same nominal final state B -- a low-T glassy state --
through two different routes in the [T,$\Gamma$] plane, presents
evidence that QA, the ``quantum route'' from A to B, yields with
the same ``cooling'' rate, a state B apparently 
closer to the ground state than CA, the classical one. 
The data however do not clarify how, and even less why, that should be so.

Theoretical suggestions and exemplifications of QA
made by various groups over the past decade
\cite{Finnila,QA_jpn,Berne1,Berne2,Farhi_Science}
have stimulated considerable interest in understanding the mechanism of
QA better. 
A theoretical discussion of the relative merits of CA and QA is therefore
desirable. 
For this, it is necessary to carry out a direct comparative test 
on a sufficiently representative benchmark system, such as a spin glass, 
and to lay the bases of a theory of the processes 
underlying QA. The issues are pressing, both because the physical underpinnings
of QA call to be explored, and because of the
practical potential of QA in the fields of optimization in complex
systems, should QA turn out to be (as recently shown in a protein
folding model\cite{Berne2}) actually superior. 
Our work is meant as a step aimed at filling these gaps.

En route, open issues are found even in the context of plain CA, where
the very rate of decay of the residual energy above the actual ground
state energy as a function of the annealing rate $1/\tau$ is controversial.
Whereas general theoretical arguments by Huse and Fisher \cite{Huse_Fisher}
predict a slow logarithmic convergence
$\epsilon_{res}(\tau)=E_{final}(\tau)-E_{GS}\sim \log^{-\zeta}(\tau)$,
with $\zeta\le 2$,
early simulations\cite{grest}, but also more recent studies \cite{CA_plaw,CA_sexp},
favor a different form, such as power-law,
$\epsilon_{res}(\tau)\sim \tau^{-\alpha}$, or stretched exponential.
The question remains whether the discrepancy
between simulations and theory is real, or only apparent.

Our work proceeded in three steps. First, we chose a benchmark
system, the Ising spin glass, where we carried out CA and QA, 
and compared the results to find that QA is indeed faster. 
Second, we focused on the residual energy in CA, to find deviations from 
power law decay versus rate $1/\tau$ that are quite compatible at 
very slow rates with the Huse-Fisher asymptotically logarithmic decay. 
Third, we built a theory of QA of a spin glass based on the idea of a cascade of
level crossings, each with its associated Landau-Zener probability
to miss the ground state. That theory suggests an asymptotic decay of residual
energy with QA rate that is again logarithmic as in CA, but governed
by somewhat different exponents that makes it faster.

Step 1. Benchmarking quantum versus classical annealing.
At the outset, inspired by Brooke et al.'s experimental system\cite{aeppli,aeppli2}, 
we selected the two-dimensional (2D) random Ising model as an appropriate 
realistic test case. 
This choice is dictated by the fact the 2D random Ising model is technically 
a polynomial problem \cite{Barahona} --
where $E_{GS}$ can be calculated up to sufficiently large lattice sizes 
\cite{germans} 
(thus avoiding an extra fitting parameter in $\epsilon_{res}(\tau)$) --
which is nonetheless of prohibitively large complexity for any physical 
dynamics as a true glass \cite{Huse_Fisher,Fisher_Huse}.  

The Edwards-Anderson Hamiltonian of an Ising spin glass in transverse field
\begin{equation} \label{eq:ising_sg}
H = - \sum_{\langle ij \rangle} J_{ij} \sigma_{i}^{z} \sigma_{j}^{z}
  -\Gamma \sum_{i} \sigma_{i}^{x} \;,
\end{equation}
where nearest-neighbor spins $\langle ij \rangle$ of a $d$-dimensional
cubic lattice interact with a random exchange coupling $J_{ij}$,
$\Gamma$ is the transverse field inducing transitions between
the two states, $\uparrow$ and $\downarrow$, of each spin, and
$\sigma_{i}^{x}, \sigma_{i}^{z}$ are Pauli matrices of the
spin 1/2 on site $i$.
The problem is to anneal this system as close as possible to its
classical, $\Gamma=0$, ground state.
In CA \cite{Kirk_CA,Cerny}, there is no tranverse field and no quantum
mechanics ($\Gamma=0$): one starts with a sufficiently 
high temperature $T_0$, which is then reduced linearly to zero in a time $\tau$.
In QA, $T$ is instead fixed to zero or some small value, and one starts
with a transverse field $\Gamma_0$ sufficiently large to throw the system
in a ``disordered'' quantum paramagnetic state, decreasing $\Gamma$ linearly to
zero, again in a time $\tau$.  
Because real time annealing is computationally out of the question for
the large systems addressed here, we carried out annealing
as a function, as customary, of the fictitious ``time'
represented by the number of Monte Carlo steps. 
Our implementation of CA was a standard Metropolis Monte Carlo (MC).  
That for QA was a path-integral Monte Carlo (PIMC) \cite{suzuki} scheme 
for a quantum system at a small finite temperature $T$.  
The 2D quantum Ising model is first mapped on a (2+1)D
classical model consisting of $P$ copies (Trotter replicas) of the original
lattice, with a nearest-neighbors uniform ferromagnetic coupling
in the third (Trotter) direction $J^{\perp}=-(PT/2) \log{ \tanh{ (\Gamma/PT)}}$,
at temperature $PT$ \cite{suzuki}.
At the beginning of the annealing, when $\Gamma$ is large, the replicas 
are only weakly coupled; as $\Gamma$ decreases to zero the ferromagnetic 
coupling $J^{\perp}$ increases, 
eventually forcing all replicas into the same configuration.
At the end of either annealing cycle the system, unable to negotiate all 
barriers in the finite time $\tau$, remains generally trapped at energy
$E_{final}=E_{GS}+\epsilon_{res}$, higher than the ground state value $E_{GS}$. 
The efficiency of each protocol is measured by the decrease of the average 
residual energy $\epsilon_{res}(\tau)$ as a function of $\tau$\cite{grest}.

For a given 2D lattice size $L\times L$, ($L$ up to 80) we took a realization
of the random couplings $J_{ij}$, drawn from a flat distribution in the
interval $(-2,2)$, and for that we got at the outset the exact classical
ground state energy $E_{GS}$ by the Branch and Cut algorithm\cite{germans}.
Keeping the couplings fixed, we then carried out a sufficient
number of repeated annealings, (45 for the $80\times 80$ lattice), 
both CA and QA. The annealing parameters
$T$ (CA) or $\Gamma$ (QA) were decreased stepwise from the initial value of
$T_{0}=3$ or $\Gamma_{0}=2.5$ down to zero, with a total of $\tau$ MC steps
per spin.   
In QA we used fixed values of $PT=1,1.5,2$ at several $P$ values,
and prepared the initial state (same for all replicas) by classical annealing 
from a temperature of 3.0 down to the corresponding value of $PT$.
In all cases the residual energy $\epsilon_{res}(\tau)$ was
calculated by subtracting $E_{GS}$ from the averaged final annealed energies.

Fig.\ 1 shows the residual energy, for both CA and QA 
for the $80\times 80$ lattice, plotted against the 
inverse annealing rate $\tau$, actually the actual Monte Carlo computer time. 
QA appears definitely superior to CA, with a lower residual energy for large $\tau$. 
This theoretical finding goes very much in the same direction
as the experimental evidence of a significantly
faster frequency-dependent relaxations
observed after QA of the disordered magnet\cite{aeppli}.
The $\tau$ dependence of our QA data does
depend on the chosen values of $P$ and $T$, particularly upon
the value of $PT$, whose optimal value appears to be around $PT=1$.
An increase of $P$ for a fixed value of $PT$, see inset in Fig.\ 1, 
ceases to be effective beyond a certain characteristic length (which
depends on $PT$) in the imaginary time direction, .
The computational cost increases linearly with $P$, and 
the choice $P=20$ (corresponding to $T=0.05$), was
found to be optimal up to the largest values of $\tau$ used.  
Another property (not shown in Fig.\ 1) of the CA results is that
residual energies obtained for different sizes $32<L<80$, or even for
different realizations of the couplings $J_{ij}$ are remarkably size
independent and self-averaging, and all fall essentially on top of the CA
curve in Fig.\ 1.

Step 2. Asymptotic behavior of classical annealing.
A feature evident in our $\epsilon_{res}(\tau)$ CA data,
is its gentle but consistent deviation from a pure power law,
suggesting  serious reconsideration of all the earlier power law 
claims \cite{grest,CA_plaw}. 
Since the slope (or apparent power) systematically declines for increasing
$\tau$, it is natural to ask whether it will asymptotically extrapolate to
zero in accordance with the Huse-Fisher logarithmic law \cite{Huse_Fisher}.
Writing that in the form  $\epsilon_{res}^{-1/\zeta}=A \log(\gamma\tau)$ and
replacing time with number of Monte Carlo steps, we can plot the CA data
as in Fig.\ 2. 
The extrapolated behavior is indeed compatible with a Huse-Fisher
straight line. However, as Fig.\ 2 shows, it proves impossible
to extract a value for the exponent $\zeta$, in particular to establish if
$\zeta\le 2$ \cite{Huse_Fisher} is any better, as one could have expected.

Step 3. Landau-Zener theory of quantum annealing.
In order to shed some light on the actual asymptotic form of 
residual energy in QA, and eventually rationalize why that might be superior,
we start off with a cartoon of the instantaneous energy spectrum of 
(\ref{eq:ising_sg}) versus $\Gamma$ in Fig.\ 3, 
suggested by small-systems exact diagonalizations. 
For sufficiently large initial $\Gamma >> |J_{ij}|$ the ground state,
generally nondegenerate\cite{mattis}, must have a finite excitation gap.  
Imagine following the Schr\"odinger evolution of an initial ground
state wavefunction $|\Psi_{\Gamma_0}(t=0)\rangle$ while
reducing $\Gamma$ gradually to zero as a function of time \cite{Farhi_Science}. 
The instantaneous gap of our disordered magnet will close as $\Gamma$ 
decreases through the quantum phase transition at $\Gamma_c$ \cite{Rieger,Thill,Wu}.
After that, ground state level crossings begin. 
The arrows in the cartoon point to two crossings 
[really {\em avoided} crossings \cite{mattis}, the problem possessing no symmetry].
Each instantaneous ground state crossing is associated with tunneling of the whole
system between two valleys -- say from a broader but shallower valley
to a narrower but deeper one, taking place when kinetic energy diminishes --
and represents a major crisis in the otherwise quasi-adiabatic evolution 
caused by the time-dependent decrease of $\Gamma(t)$.  

For sufficiently slow annealing, each tunneling event can be treated 
as a Landau-Zener (LZ) problem\cite{Landau,Zener}, see inset in Fig.\ 3.  
The probability $P(\tau)$ that the system, starting in the lower state 
$|b\rangle$ at high $\Gamma$ will continue nonadiabatically onto the higher 
branch as $\Gamma$ is reduced with time is given by 
$P(\tau)=\exp{(-\tau/\tau_c)}$ where $\tau_c$, the characteristic 
{\em tunneling time}, is 
$\tau_c=(\hbar \alpha \Gamma_0)/(2\pi\Delta^2)$.
Here $\Delta$ is the tunneling amplitude between the two states
$|a\rangle$ and $|b\rangle$ (whose splitting at crossing is 2$|\Delta|$),
and $\alpha$ is the relative slope of the two crossing branches as
a function of $\Gamma$ \cite{Landau,Zener}. 
One can estimate $\Delta \sim e^{-d_{ab}/\xi(\Gamma)}$, where $d_{ab}$ is
a suitable distance between states $a$ and $b$ (in the Ising case, the number
of spins that are flipped in the tunneling process, $N_{flip}$\cite{Thill}), 
and $\xi(\Gamma)$ is a typical wavefunction localization length,
which must vanish as $\Gamma\to 0$, $\xi(\Gamma)\sim \Gamma^\phi$,
with some exponent $\phi>0$.
The tunneling time becomes exponentially large
for small $\Gamma$, $\tau_{\Gamma} \sim e^{2d_{ab}/\Gamma^\phi}$,
and an exceedingly small width $\sim \Delta$ of each tunneling event
justifies treating the multiple crossings as a cascade of independent
LZ events.
Once the system fails, with a probability
$P_{\Gamma}(\tau)=e^{-\tau/\tau_{\Gamma}}$, to follow the ground state at the LZ
crossing occurring at $\Gamma$, it will eventually attain
an average excitation energy $E_{ex}(\Gamma)>0$.
Letting $Z(\Gamma)d\Gamma$ be the number of LZ crossings which take place
between $\Gamma$ and $\Gamma+d\Gamma$, the average residual energy can be
estimated as
\begin{equation}
\epsilon_{res}(\tau)
= \int_0^{\Gamma_c} d\Gamma \; Z(\Gamma) E_{ex}(\Gamma) \;
e^{-\tau/\tau_{\Gamma}} \;,
\end{equation}
where $\Gamma_c$ marks the first level crossing.
The large $\tau$ behavior of this expression is dominated by the 
$\Gamma\to 0$ behavior of $Z(\Gamma) E_{ex}(\Gamma)$, and $\tau_{\Gamma}$.  
If we assume that, for small $\Gamma$, 
$Z(\Gamma) E_{ex}(\Gamma)\sim \Gamma^{\omega}$, and 
$\tau_{\Gamma}\sim e^{A/\Gamma^\phi}$ we get finally a residual energy 
which vanishes as the inverse power of the rate logarithm
$\epsilon_{res}(\tau) \sim \log^{-\zeta_{QA}}(\tau)$,
with an exponent $\zeta_{QA}=(1+\omega)/\phi$.
The exponents $\omega$ and $\phi$ are not obvious.
A semiclassical (WKB) expression for the decay of a wavefunction inside a barrier
suggests $\phi=1/2$.
The average excitation energy attained by
missing the ground state ``track'' at $\Gamma$ should scale as $\Gamma^2$
for small $\Gamma$, because all eigenvalues start out as $\Gamma^2$ for
$\Gamma\to 0$.
The total number of LZ crossings occurring from $0$ to $\Gamma$ should not be
larger than the total number of classical 
states in the energy window $(E_{GS},E_{GS}+\Gamma)$, which is 
approximately equal to 
$\rho(0) \Gamma$ [where $\rho(0)$ s the density of classical states
at the ground state energy \cite{Fisher_Huse}], 
so that the density of crossings $Z(\Gamma\to 0)\to \rho(0)$, at most. 
This yields $\omega=2$ as our most reasonable estimate.  

We conclude that $\zeta_{QA}=(1+\omega)/\phi$ can be as
large as $6$ for a spin glass, and in any case above the classical 
Huse-Fisher bound $\zeta\le 2$ \cite{Huse_Fisher}.  
Hence, quantum annealing of the Ising spin glass is predicted to be again 
logarithmically accurate, 
not fundamentally different in that from classical annealing.
We therefore expect that a quantum computation based on QA will not transform 
a hard nonpolynomial (NP-complete) computational problem into a polynomial one. 
On the contrary, the above reasoning suggests a logarithmically 
slow annealing to apply also to the present 2D Ising case, which is not 
NP-complete \cite{Barahona}. 
The slowing down effect of the LZ cascade 
illustrated above is particularly severe in problems, like the 
Ising spin glass we have considered, where the classical spectrum has a 
{\em gapless} continuum of excitations above the ground state.
Satisfiability problems, for which much more encouraging results were
recently presented \cite{Farhi_Science} differ from the Ising spin glass in 
that they possess a
discrete classical spectrum and a finite excitation gap. 
We observe that in general a gap will cut off the LZ cascade
precisely in the dangerous low-$\Gamma$ region, and that may eliminate 
the logarithmic slowing down of QA.
Nonetheless, even in the gapless case, the advantage of QA over CA
is far from negligible, due to the generally larger exponent $\zeta_{QA}$ 
of the logarithm.
To get an idea of the order of magnitudes involved, consider 
the relative increase of annealing time $(\tau'/\tau)$ needed to improve the 
accuracy of a certain annealing, say with $\tau \sim 10^6$ 
(in appropriate units) by a factor 10.
In CA ($\zeta=2$), this would require 
$(\tau'/\tau)\sim \tau^{10^{1/\zeta}-1}\sim 10^{13}$.
In QA ($\zeta=6$), the same result would be accomplished with
$(\tau'/\tau)\sim 10^{2.8}$, an enormous saving of computer effort.  
Moreover, the PIMC version of QA is easy to implement on a parallel computer, 
and that provides an extra advantage.

In summary, our test of QA in the disordered Ising magnet indicates a
faster convergence than CA, and a time-dependent cascade of Landau-Zener 
tunneling events across barriers is pinpointed as the crucial ingredient of QA. 
Optimization by QA of a vast variety of problems beyond statistical
mechanics, of course after a suitable fictitious kinetic energy operator is 
identified case by case, is an open avenue, and stands as a worthy 
challenge for the future.


This project was sponsored by MIUR under project COFIN, by INFM/F, INFM/G,
and by INFM's ``Iniziativa Trasversale Calcolo Parallelo''.
R.M acknowledges European Union support through CINECA under project MINOS3,
which also provided much of the computer resources.
We are grateful to G. Aeppli, L. Arrachea, J. Berg, C. Micheletti, 
M. Parrinello, F. Ricci Tersenghi, R. Zecchina, 
for helpful discussions and suggestions. 


%
\newpage
\begin{figure}
\centerline{\psfig{figure=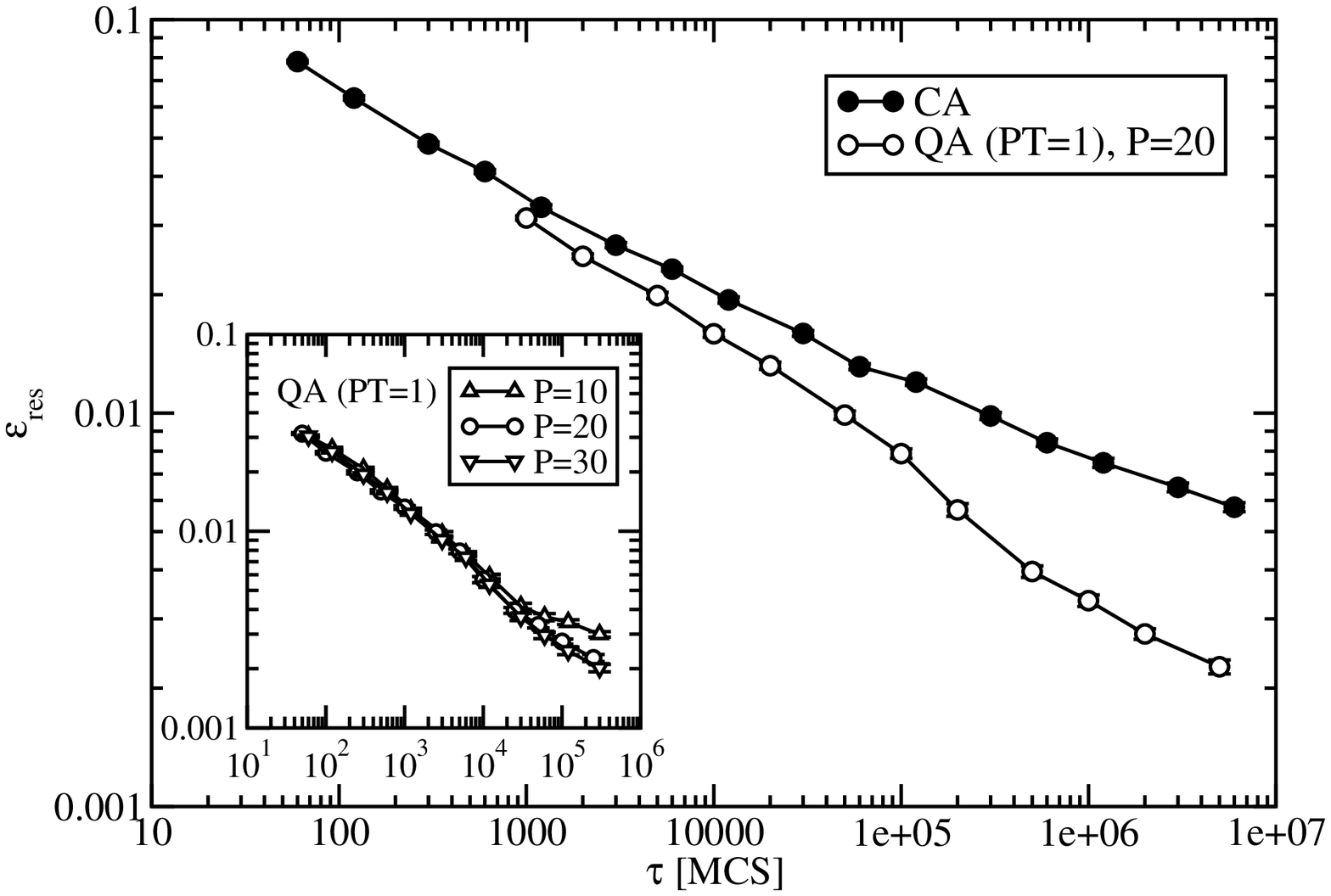,height=20cm}}
\caption{ 
  Comparison of the residual energy per site for an $80\times 80$ 
  disordered 2D Ising model after classical and quantum annealing.
  The QA data shown correspond to the optimal value
  of $PT=1$, with $T=0.05$ and $P=20$ Trotter replicas. 
  For fair comparison, the actual inverse annealing rate $\tau$ used in 
  the QA has been rescaled (multiplied by $P$) so that points at the same $\tau$ 
  require the same computer time (MCS, Monte Carlo steps). 
  The lower residual energy signifies that QA is superior to CA.  
  Inset: $\tau$-unrescaled QA data for the same system for 
  increasing values of $P$. Note the satisfactory convergence for $P=20$.}
\end{figure}

\newpage
\begin{figure}
\centerline{\psfig{figure=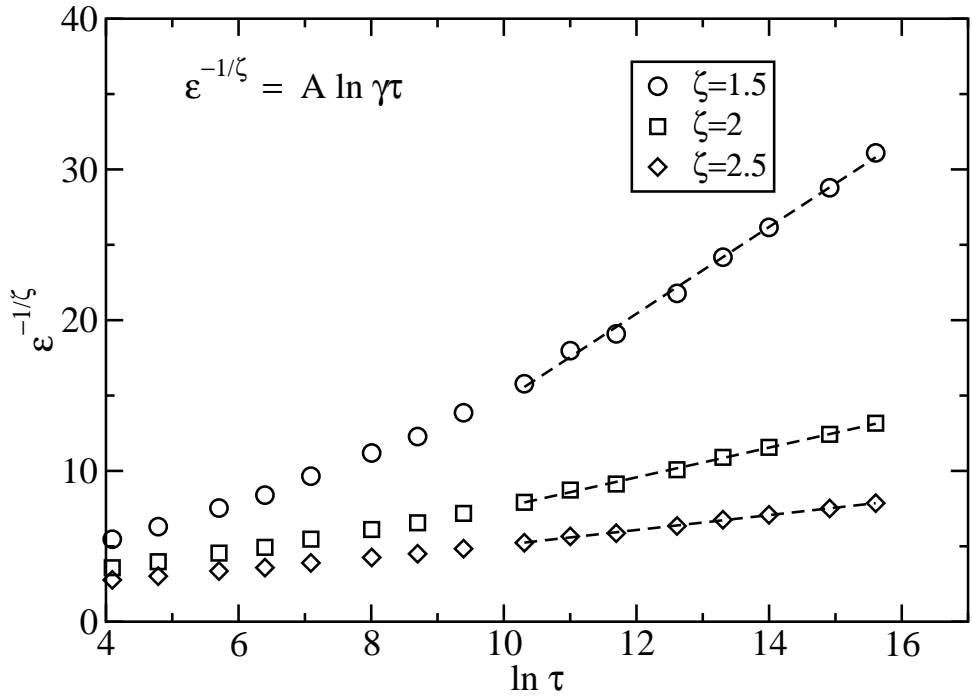,height=20cm}}
\caption{
The same CA data as in Fig.\ 1 re-plotted (see text) so as to fall on a
straight line if obeying the Huse-Fisher logarithmic law. Although the Huse-Fisher
form is seen to be asymptotically compatible with the data, extraction of a
value for the exponent $\zeta$ is impossible.
}
\end{figure}

\newpage
\begin{figure}
\centerline{\psfig{figure=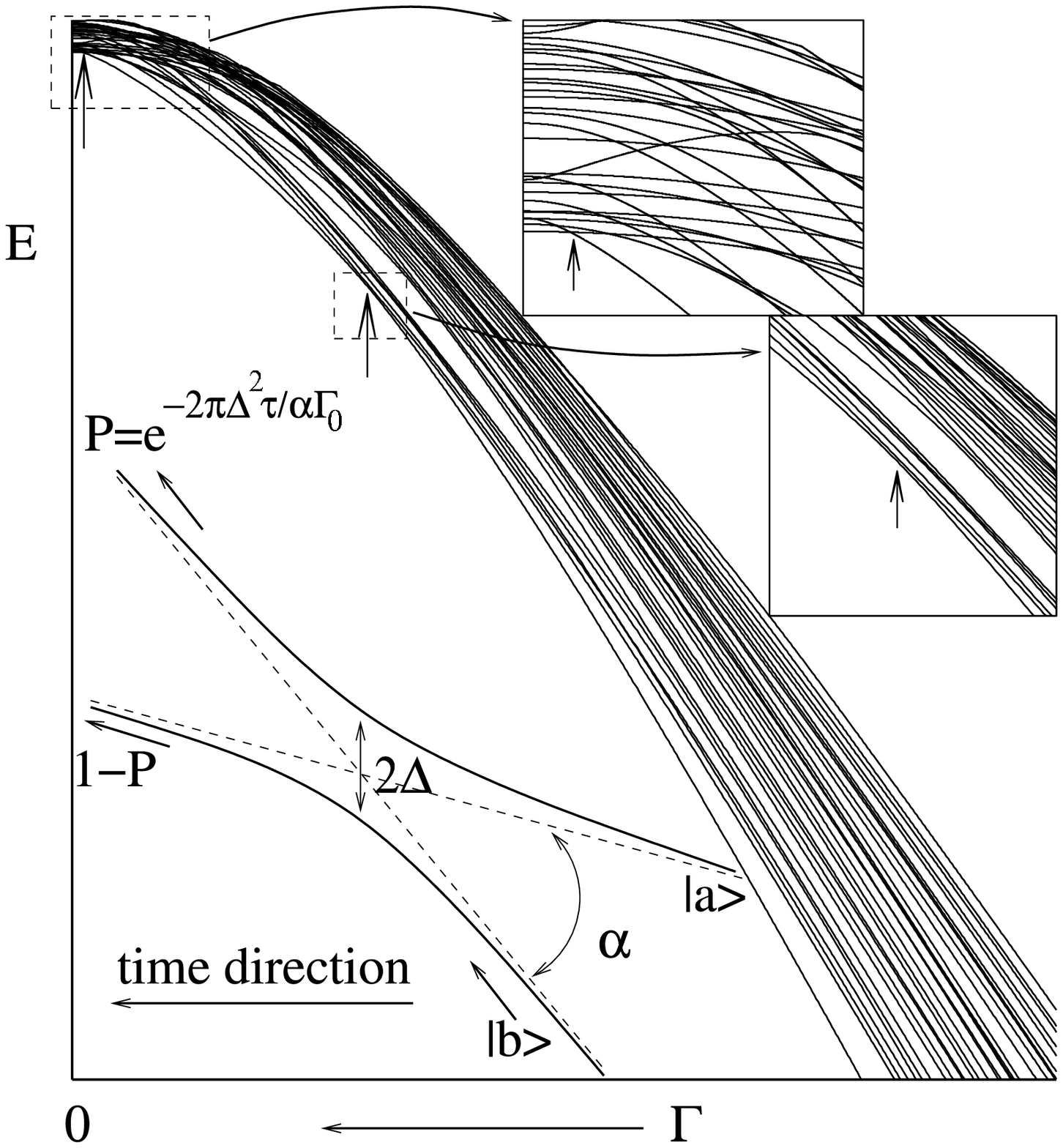,height=15cm}}
\caption{
Cartoon of the lowest instantaneous eigenvalues of a (finite-size)
Ising glass as a function of the transverse field $\Gamma$, or
of a generic complex system as a function of its zero-point
kinetic energy $\Gamma$.
Note the two avoided crossing of the ground state, marked by arrows and 
enlarged in the upper insets. 
Lower inset: Schematic of a Landau-Zener crossing.
At each crossing the system will follow adiabatically the ground state 
only if $\Gamma$ is reduced sufficiently slowly. The infinite system
will exhibit an infinite cascade of crossings as $\Gamma\to 0$.
}
\end{figure}


\begin{thebibliography}{99}

\bibitem{Kirk_CA}
S. Kirkpatrick, C.D. Gelatt, Jr., M.P. Vecchi, Science {\bf 220}, 671 (1983).

\bibitem{Cerny}
V. \v{C}ern\'y, J. Opt. Theor. Appl. \textbf{45}, 41 (1985).

\bibitem{Huse_Fisher}
D.A. Huse, D.S. Fisher, Phys.\ Rev.\ Lett.\ {57}, 2203 (1986).

\bibitem{aeppli}
J. Brooke, D. Bitko, T.F. Rosenbaum, G. Aeppli,
Science {\bf 284}, 779 (1999).

\bibitem{aeppli2}
J. Brooke, T.F. Rosenbaum, G. Aeppli, Nature {\bf 413}, 610 (2001).

\bibitem{Finnila}
A.B. Finnila, M.A. Gomez, C. Sebenik, C. Stenson, J.D. Doll,
Chem.\ Phys.\ Lett.\ {\bf 219}, 343 (1994).

\bibitem{QA_jpn}
T. Kadowaki, H. Nishimori, Phys.\ Rev.\ E {\bf 58}, 5355 (1998).

\bibitem{Berne1}
Y.H. Lee, B.J. Berne, J. Phys.\ Chem.\ A {\bf 104}, 86 (2000).

\bibitem{Berne2}
Y.H. Lee, B.J. Berne, J. Phys.\ Chem.\ A {\bf 105}, 459 (2001).

\bibitem{Farhi_Science}
E. Farhi, J. Goldstone, S. Gutmann, J. Lapan, A. Lundgren, D. Preda,
Science {\bf 292}, 472 (2001).  

\bibitem{grest}
G.S. Grest, C.M. Soukoulis, K. Levin, Phys.\ Rev.\ Lett.\ {56}, 1148 (1986).

\bibitem{CA_plaw}
A. Chakrabarti, R. Toral, Phys.\ Rev.\ B {\bf 39} 542 (1989).

\bibitem{CA_sexp}
P. Ocampo-Alfaro, H. Guo, Phys.\ Rev.\ B {\bf 53}, 1982 (1996).

\bibitem{Barahona}
F. Barahona, J. Phys.\ A {\bf 15}, 3241 (1982).

\bibitem{germans}
C. De Simone {\it et al.}, J. Stat.\ Phys.\ {\bf 80}, 487 (1995).
For each set of couplings, $E_{GS}$ is calculated using the Spin
Glass Ground State Server at
www.informatik.uni-koeln.de/ls\_juenger/projects/sgs.html. 

\bibitem{Fisher_Huse}
D.S. Fisher, D.A. Huse, Phys.\ Rev.\ Lett.\ {56}, 1601 (1986).

\bibitem{suzuki}
M. Suzuki, Prog.\ Theor.\ Phys.\ \textbf{56}, 1454 (1976).
See also
\textit{Quantum Monte Carlo Methods in Equilibrium and Nonequilibrium
Systems}, Proceedings of the Ninth Taniguchi International Symposium,
Susono, Japan, 1986, ed. by M. Suzuki (Springer-Verlag Berlin Heidelberg
1987)

\bibitem{mattis} 
E. Lieb, D. Mattis, J. Math.\ Phys.\ {\bf 3}, 749 (1962). 

\bibitem{Rieger}
H. Rieger, A.P. Young, Phys.\ Rev.\ Lett.\ {\bf 72}, 4141 (1994). 

\bibitem{Thill}
M.J. Thill, D.A. Huse, Physica A {\bf 214}, 321 (1995).

\bibitem{Wu}
W. Wu, D. Bitko, T.F.Rosenbaum, G. Aeppli, Phys.\ Rev.\ Lett.\ {\bf 71}, 1919 
(1993).

\bibitem{Landau}
L.D. Landau, Phys.\ Z. Sowjetunion {\bf 2}, 46 (1932).

\bibitem{Zener}
C. Zener, Proc.\ Royal Soc.\ A {\bf 137}, 696 (1932).

\end{thebibliography}
\end{document}